\documentclass[twocolumn,showpacs,preprintnumbers,amsmath,amssymb,prb]
{revtex4}

\usepackage{amsmath,amssymb}
\usepackage{graphicx}
\usepackage{dcolumn}
\usepackage{bm}
\usepackage[dvips]{color}

\def\rnum#1{\expandafter{%
\romannumeral #1}}
\def\Rnum#1{\uppercase\expandafter{%
\romannumeral #1}}

\begin{document}


\title{NMR relaxation rate and dynamical structure factors
in nematic and multipolar liquids
of frustrated spin chains under magnetic fields}

\author{Masahiro Sato}
\affiliation{Condensed Matter Theory Laboratory, RIKEN, Wako, Saitama
351-0198, Japan}
\author{Tsutomu Momoi}
\affiliation{Condensed Matter Theory Laboratory, RIKEN, Wako, Saitama
351-0198, Japan}
\author{Akira Furusaki}
\affiliation{Condensed Matter Theory Laboratory, RIKEN, Wako, Saitama
351-0198, Japan}



\date{\today}

\begin{abstract}
Recently, it has been shown that spin nematic (quadrupolar) or
higher multipolar correlation functions
exhibit a quasi long-range order in the wide region of the field-induced
Tomonaga-Luttinger-liquid (TLL) phase in spin-$\frac{1}{2}$ zigzag
chains. In this Rapid Communication,
we point out that the temperature dependence of
the NMR relaxation rate $1/T_1$ in these multipolar TLLs is qualitatively
different from that in more conventional TLLs of
one-dimensional quantum magnets (e.g., the spin-$\frac12$ Heisenberg chain);
$1/T_1$ decreases with lowering temperature in multipolar TLL. 
We also discuss low-energy features
in spin dynamical structure factors
which are characteristic of the multipolar TLL phases.
\end{abstract}

\pacs{76.60.-k,75.40.Gb,75.10.Jm,75.10.Pq}

\maketitle

Magnetic states with an order parameter defined by
a product of multiple spins, such as nematic,
vector chiral, and scalar chiral orders,
have attracted much attention.
Spin nematic (quadrupolar) ordered phases have been recently shown
to appear in frustrated ferromagnets,
such as ferromagnets with competing antiferromagnetic (AF)
interactions\cite{Chubukov,Momoi1,Momoi2} and
magnets with multi-spin-exchange
couplings.\cite{Tsunetsugu,Pen,Momoi1}
A triatic (octupolar) ordered
phase was also found in the triangular lattice multiple-spin exchange
model with ferromagnetic (FM) dominant coupling.\cite{Momoi2}
These nematic and triatic ordered states can be regarded as
Bose condensed states of bound two-magnons\cite{Chubukov,Momoi1}
and bound three-magnons,\cite{Momoi2} respectively, and
their order parameters are given by
\label{nematic}
$S^+_j S^+_k$ or $S^-_j S^-_k$, and 
$S^+_j S^+_k S^+_l$ or $S^-_j S^-_k S^-_l$.

Recent extensive studies~\cite{Kolezhuk,Vekua,Kecke,Hikihara1,Sudan} 
have shown that a series of similar multipolar phases
appear in the one-dimensional (1D) spin-$\frac{1}{2}$ Heisenberg model
with FM nearest-neighbor exchange $J_1$ and competing AF
next-nearest-neighbor exchange $J_2$
in applied field $H$, whose Hamiltonian is
\begin{equation}
\label{zigzag}
{\cal H}= \sum_{n=1,2} \sum_{j} J_n \bm{S}_j\cdot\bm{S}_{j+n}
-H\sum_{j} S^z_j.
\end{equation}
Here $\bm{S}_j$ is the spin-$\frac{1}{2}$ operator on $j$th site,
$J_1<0$ and $J_2>0$, and $H$ is the external magnetic field in the 
$z$ direction. This simple frustrated spin chain is a minimal model of
frustrated ferromagnets and is thought to describe magnetism
in quasi-1D edge-sharing
cuprates,~\cite{Hase,Drechsler,Enderle-Naito,Masuda-Park-Seki}
such as $\rm Rb_2Cu_2Mo_3O_{12}$, $\rm NaCu_2O_2$, $\rm LiCuVO_4$, and
$\rm LiCu_2O_2$.

Hikihara \textit{et al}.\cite{Hikihara1}\
and Sudan {\it et al}.\cite{Sudan}\ showed
that the ground state of Hamiltonian (\ref{zigzag}) has
field-induced Tomonaga-Luttinger (TL) liquid phases
in which
spin multipolar correlations are
quasi-long-range ordered while the transverse spin correlation
is short-ranged.
This result can be easily understood in the large-magnetization regime,
where $p$ magnons form a bound state ($p>1$).~\cite{Kecke,Hikihara1}
A gas of bound $p$ magnons acquires
off-diagonal quasi-long-range order, with the order parameter
being the effective hard-core boson creation operator
$\prod_{l=1,\cdots,p}S_{j+l}^-$.
In the original spin language this off-diagonal correlation is
the multipolar spin correlation characterizing
a nematic ($p=2$), an octupolar ($p=3$), or a hexadecapolar
($p=4$) phase.
Numerical studies~\cite{Kecke,Hikihara1,Sudan} found $p=2$ for
$-2.7\alt J_1/J_2<0$, $p=3$ for $-3.5\alt J_1/J_2\alt-2.7$,
and $p=4$ for $-3.76\alt J_1/J_2\alt-3.5$, at the
saturation field.
At lower magnetic fields these phases cross over to
spin-density wave SDW$_p$ phases in which
the density correlation of bound $p$ magnons,
i.e., the longitudinal SDW, becomes stronger
than the multipolar correlation.
Incidentally, a SDW$_2$ phase is present also
in the case of AF $J_1>0$.\cite{Hikihara2,Okunishi}

However, it will be difficult to
obtain direct experimental evidence for the multipolar spin orders,
as it requires probing four- or more-spin correlation
functions with high accuracy.
Standard experimental probes, such as neutron scattering or 
magnetic resonance, measure only two-spin correlations.
Furthermore, the multipolar TL liquids
have a gapless spectrum and a smooth magnetization curve.
Thus, if one only measures their static,
thermodynamic quantities 
(uniform susceptibility, specific heat, entropy, etc.),
it is hard to distinguish the multipolar TL liquids from
conventional TL liquids.
Experimental schemes for identifying multipolar spin orders
are therefore called for.

\begin{figure}[tth]
\begin{center}
\includegraphics[width=8cm]{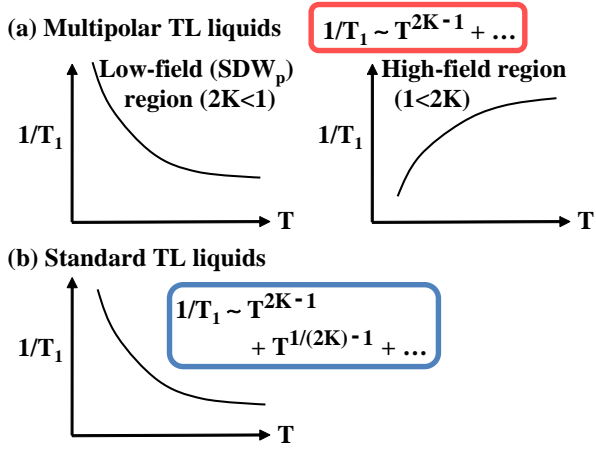}
\end{center}
\caption{(color online) Temperature and field dependence of
the NMR relaxation rate $1/T_1$ in (a) multipolar phases
in spin-$\frac{1}{2}$ frustrated zigzag chains
and in (b) standard TL liquids
(e.g., the AF Heisenberg chain in magnetic field).
The parameter $K$ controls correlations in TL liquids (see the text).
}
\label{NMR_rate}
\end{figure}


In this Rapid Communcation, 
we propose that NMR measurements can capture signatures 
(albeit indirect) of the multipolar TL liquids. We show that,
in the TL liquids with a dominant multipolar spin correlation,
the NMR relaxation rate $1/T_1$ decreases
as temperature $T$ is lowered [see Fig.~\ref{NMR_rate}(a)].
This temperature dependence of $1/T_1$ is opposite to that
in conventional TL liquids (and in SDW$_p$ phases),
where it is always diverging as $T\to0$,
in magnetic field.\cite{Giamarchi} We also point out
that  the spin dynamical structure factors exhibit
features which are very characteristic of the multipolar TL liquids.

Let us begin with a brief review of
the effective theories~\cite{Vekua,Hikihara1}
for the multipolar TL liquids, which allow us to find
low-energy behavior of spin correlation functions
observed in NMR and in dynamical structure factors.
In the weak $J_1$ limit, the Abelian bosonization method~\cite{Gia_text}
is useful. It leads to nematic ($p=2$) and 
SDW$_2$ phases~\cite{Kolezhuk,Vekua,Hikihara1} and
a vector chiral ordered phase.~\cite{Nersesyan} The effective
Hamiltonian for the nematic ($p=2$) and SDW$_2$
phases~\cite{Kolezhuk,Vekua,Hikihara1} is written as
\begin{eqnarray}
\label{zigzag_eff}
{\cal H}_{\rm eff} \!\!&=&\!\! \int\, dx
\Big\{\sum_{\nu=\pm}\frac{v_\nu}{2}
\left[K_\nu^{-1}(\partial_x\phi_\nu)^2+K_\nu(\partial_x\theta_\nu)^2\right]
\nonumber\\
&&\hspace*{10mm}{}
+g\sin(\pi M)\sin(\sqrt{8\pi}\phi_-+\pi M)\Big\},
\end{eqnarray}
where $x=2j$ (the lattice spacing is set equal to unity),
$(\phi_\pm,\theta_\pm)$ is a pair of dual scalar fields
satisfying the commutator
$[\phi_\mu(x), \partial_y\theta_\nu(y)]=i\delta_{\mu,\nu}\delta(x-y)$,
$M=\langle S_j^z\rangle$, $g\propto J_1$,
and $K_\pm$ and $v_\pm$ are, respectively, the TL-liquid
parameter and the velocity
of the $(\phi_\pm,\theta_\pm)$ sector.
The $(\phi_+,\theta_+)$
sector is a gapless TL liquid,
while the $(\phi_-,\theta_-)$ sector has a gapful spectrum
because the field $\phi_-$ is pinned at a value minimizing the
potential energy of the $g$ term.
Using Hamiltonian~(\ref{zigzag_eff}),
one can evaluate the low-energy and long-distance behaviors of several
correlation functions. The imaginary-time ($\tau$) spin and nematic
correlations at zero temperature $T=0$ are calculated as
\begin{subequations}
\label{correlators_nematic}
\begin{align}
\label{correlators_nematic_1}
\!\!
&\langle S^z_j(\tau)S^z_0(0)\rangle
=
M^2 - \frac{K_+}{2\pi^2}
\left(\frac{1}{z_+^2}+\frac{1}{\bar z_+^2}\right)
\nonumber\\
&\hspace*{2.2cm}
+\frac{C_1}{|z_+|^{K_+}}
\cos\!\left[\pi j\!\left(M+\frac{\mathrm{sgn}(J_1)}{2}\right)\right]
\!\!
+\cdots,\nonumber\\
\\
\label{correlators_nematic_2}
&\langle S^+_j(\tau)S^-_0(0)\rangle
=
C_2\cos \Big(\frac{\pi j}{2}\Big)
\frac{e^{-|z_-|/\xi}}{|z_+|^{1/(4K_+)}|z_-|^{1/2}}
+\cdots,
\\
\label{correlators_nematic_3}
&\langle S^+_j(\tau)S^+_{j+1}(\tau)S^-_0(0)S^-_1(0)\rangle
=
C_3\frac{(-1)^j}{|z_+|^{1/K_+}}
+\cdots,
\end{align}
\end{subequations}
where $z_\pm=j-iv_\pm\tau$,
and $C_n$ are nonuniversal positive constants.
The exponential decay of the transverse spin correlation
in Eq.\ (\ref{correlators_nematic_2}) is
qualitatively 
different from a power-law decay form in ordinary TL liquids
(e.g., the spin-$\frac12$ AF chain and ladder in magnetic field).
The correlation length $\xi$
is inversely proportional to the gap of the $(\phi_-,\theta_-)$ sector.

More generally, the TL-liquid behavior in
\textit{all} the multipolar and SDW$_p$ phases ($p \ge 2$)
can be understood from a hard-core Bose gas picture
of bound $p$ magnons,
when the nearest-neighbor coupling $J_1$ is
ferromagnetic.\cite{Kecke,Hikihara1}
Below the saturation field a (dilute) Bose gas of bound $p$ magnons
form a TL liquid with off-diagonal quasi-long-range order,
i.e., $p$th multipolar TL liquid.
In this picture one may replace the $p$th multipolar operator
$S_{j+1}^-S_{j+2}^-\cdots S_{j+p}^-$ and magnon density $\frac{1}{2}-S^z_j$
with a creation operator
of a hard-core boson $(-1)^j b_j^\dag$ and boson density
$pb_j^\dag b_j^{}$,
respectively.
Here the staggered factor $(-1)^j$ represents the total momentum
$k=\pi$ of the lowest-energy bound states.
The hydrodynamic theory for the bosonic TL liquid
has the same form as the free boson Hamiltonian of the $(\phi_+,\theta_+)$
sector in Eq.\ (\ref{zigzag_eff}).
The effective theory gives
the following longitudinal spin and the multipolar correlation functions
at $T=0$:
\begin{subequations}
\label{correlators_boson}
\begin{eqnarray}
\label{correlators_boson_1}
&&
\langle S^z_j(\tau)S^z_0(0)\rangle
=
M^2 - \frac{p^2 K}{4\pi^2}
\left(\frac{1}{z^2}+\frac{1}{\bar z^2}\right)
\nonumber\\
&&\hspace{23mm}{}
+\frac{C_4 p^2}{|z|^{2K}}\cos\!\left[{\frac{\pi j}{p}(1-2M)}\right]
\!
+\cdots,
\;\;\quad
\\
\label{correlators_boson_2}
&&
\Big\langle\prod_{n=1}^{p}S^+_{j+n}(\tau)\prod_{n=1}^{p}S^-_{n}(0)
\Big\rangle
=
C_5\frac{(-1)^j}{|z|^{1/(2K)}}
+\cdots,
\end{eqnarray}
\end{subequations}
where $z=j-iv \tau$, $K$ is the TL-liquid parameter
for the hard-core bosons.
While we cannot evaluate the transverse spin correlations
within this boson picture, they must
decay exponentially $\propto \exp(-|z|/\xi)$ as it is necessary to
break a magnon bound state in order to create an excitation
with $\Delta S^z=\pm 1$. In the nematic case of $p=2$ and $J_1<0$,
Eq.~(\ref{correlators_boson}) coincides with
Eqs.~(\ref{correlators_nematic_1}) and (\ref{correlators_nematic_3})
if we set $K_+=2K$ and $v_+=v$.
Near the saturation field where the density of magnons vanishes,
the value of $K$ approaches unity, i.e.,
that of the 1D free fermions.
Indeed, the numerical calculations in
Ref.~\onlinecite{Hikihara1} have shown that $K$ monotonically
increases from about $1/4$ to unity with the increase in the
magnetization $M$.
This means that the multipolar correlation (\ref{correlators_boson_2})
is strongest in the high-field regime ($2K>1$),
whereas the SDW correlation (\ref{correlators_boson_1})
becomes most dominant in the low-field regime ($2K<1$).
This property is important in the following discussion
on the NMR relaxation rate.
We note that at $p=1$, Eq.~(\ref{correlators_boson}) reproduces
the spin correlations in the TL-liquid phase of, e.g., the
spin-$\frac{1}{2}$ AF chains under magnetic field.~\cite{Gia_text}

The temperature dependence of the NMR relaxation rate $1/T_1$ in the
multipolar TL liquids can be derived from
the above asymptotic forms of correlation functions.
The perturbation theory in hyperfine interaction between nuclear
and electron spins obtains $1/T_1$ as~\cite{Giamarchi,Goto}
\begin{eqnarray}
\label{NMR_formula}
\frac{1}{T_1}\!\!&\propto&\!\!
\sum_k \biggl\{\frac{\left|A_k^\perp\right|^2}{2}
\left[S^{+-}(k,\omega)+S^{-+}(k,\omega)\right]
\nonumber\\
&&\hspace*{8mm}
+\bigl|A_k^\parallel\bigr|^2S^{zz}(k,\omega)\biggr\},
\end{eqnarray}
where $\omega$ is the nuclear resonance frequency,
$A_k^\nu$ are the hyperfine form factors,
and $S^{\alpha\beta}(k,\omega)=\sum_{j} e^{-ikj}
\int_{-\infty}^\infty dt \,
e^{i\omega t}\langle S_j^\alpha(t)S_0^\beta(0)\rangle$
is the spin dynamical structure factor ($t=-i\tau$ is the real time) at
temperature $T$.
Since $\omega$ is generally much smaller than the energy scale of
spin exchange interactions, we may take the limit $\omega/T\to+0$.
Moreover, the $k$ dependence of $A_k^\nu$ is usually weak due to the
locality of the nucleus-electron interaction.
Hence, the $T$ dependence of $1/T_1$ can be obtained by
evaluating the local susceptibility,
$\int_{-\infty}^\infty dt \,
e^{i\omega t}\langle S_j^\alpha(t)S_j^\beta(0)\rangle$.

The local susceptibility
at finite temperatures
can be readily obtained from
the correlation functions
(\ref{correlators_nematic}) and
(\ref{correlators_boson})
through the standard procedure.\cite{Gia_text,Giamarchi}
Substituting them into Eq.\ (\ref{NMR_formula}), we obtain
$1/T_1$ for the multipolar TL liquids in the form
\begin{equation}
\label{T1_multipolar}
1/T_1 =  D_1^\parallel T + D_2^\parallel T^{2K-1}+\cdots.
\end{equation}
The two leading terms
come from the second and third terms,
respectively,
of the longitudinal spin correlation,
Eqs.~(\ref{correlators_nematic_1}) or (\ref{correlators_boson_1}).
The coefficients $D_1^\parallel$ and $D_2^\parallel$
are independent of temperature in the regime
$\omega\ll T\ll |J_{1,2}|$.
In Eq.~(\ref{T1_multipolar}) we have omitted
contributions from the transverse spin correlations
which are exponentially small,
$e^{-\Delta/T}$ ($\Delta\approx v/\xi$ is proportional to
the spin gap),
at low temperatures $T\ll |J_{1,2}|$.
When $K<1$, the second term in Eq.\ (\ref{T1_multipolar})
gives the leading contribution in the low-temperature limit.
Similarly, the known $T$ dependence of
$1/T_1$ in spin-$\frac{1}{2}$ AF chains
under a magnetic field is obtained from
Eq.\ (\ref{correlators_boson}) with $p=1$ in the
form~\cite{Giamarchi,Gia_text}
\begin{equation}
\label{T1_TLL}
1/T_1 = E_1^\parallel T + E_2^\parallel T^{2K-1}
+E_1^\perp T^{1/(2K)-1}+\cdots,
\end{equation}
where the terms $\propto E_n^\parallel$ and $E_n^\perp$
are derived
from the longitudinal and the transverse spin correlations,
respectively. Equation (\ref{T1_TLL}) commonly holds in TL-liquid phases
of 1D magnets such as AF spin chains and
ladders in magnetic field.~\cite{Giamarchi}

Comparison of Eqs.~(\ref{T1_multipolar}) and (\ref{T1_TLL})
tells us
an important feature of the NMR relaxation rate in the multipolar TL
liquids.
As we noted above,
the parameter $K$ in the multipolar phases
of Hamiltonian (\ref{zigzag}) with the FM coupling $J_1<0$,
is an increasing function of $H$ and approaches unity
at the saturation field.\cite{Hikihara1}
The monotonic magnetic-field dependence of $K$ presumably
holds for other multipolar TL liquids as well,
at least for spin-$\frac12$ AF spin systems.
Equation (\ref{T1_multipolar}) then implies that
$1/T_1$ decreases with lowering temperature in
the high-field multipolar phase ($2K>1$),~\cite{Note2}
while it shows diverging behavior
in the low-field SDW$_p$ region ($2K<1$); see Fig.~\ref{NMR_rate}.
This behavior is totally different from that of 
conventional TL liquids like
AF spin chains under magnetic field [Eq.\ (\ref{T1_TLL})],
in which $1/T_1$ always diverges
in the low-temperature limit,
irrespective of the value of $K$ (the case of $K=1/2$ is
special~\cite{Takigawa,Sachdev}).
We emphasize that this difference in the $T$ dependence of $1/T_1$
between multipolar and conventional TL liquids,
shown in Figs.~\ref{NMR_rate} (a) and (b), can be
taken as a pronounced signature of 1D spin-$\frac{1}{2}$
multipolar TL-liquid phases.
The decay of $1/T_1$ with lowering temperature
in the multipolar liquid phases is due to both
the absence of gapless modes in $S^{+-}(k,\omega)$ and
the weak singularity at $\omega=0$ in $S^{zz}(k,\omega)$
[see Eqs.~(\ref{Sqw_nematic_FM}) and (\ref{Sqw_multipolar})].
We also note that NMR experiments cannot distinguish a SDW$_p$ region
from ordinary TL liquids
because they both show divergent behavior of $1/T_1$ as $T\to0$.

Next we discuss the spin dynamical structure factors
$S^{\alpha\beta}(k,\omega)$ at $T=0$ in the multipolar phases.
The support of $S^{\alpha\beta}(k,\omega)$ tells us
which excitations in the $(k,\omega)$ space contribute to
inelastic neutron scattering.
The low-energy parts of $S^{\alpha\beta}(k,\omega)$
are obtained from
Fourier transform~\cite{Furusaki} of
correlation functions~(\ref{correlators_nematic}) and
(\ref{correlators_boson}).
For the nematic and SDW$_2$ phases in FM $J_1<0$,
we find
\begin{subequations}
\label{Sqw_nematic_FM}
\begin{eqnarray}
\label{Sqw_nematic_FM1}
&&
S^{zz}(k\sim 0,\omega) = 4K|k|\delta(\omega-v|k|),\\
\label{Sqw_nematic_FM2}
&&
S^{zz}(k\sim \pm k_2,\omega) =
\frac{c^z_2\Theta_s(\omega-v|k\mp k_2|)}
{[\omega^2+v^2(k\mp k_2)^2]^{1-K}},\\
&& \!\!
S^{+-}(k\sim \pm \pi/2,\omega) =
\frac{c^\perp_2\Theta_s\biglb(\omega-\epsilon(k\mp\pi/2)\bigrb)}
{[\omega-\epsilon(k\mp\pi/2)]^{1-1/(4K_+)}},
\;\quad
\end{eqnarray}
\end{subequations}
where $k_2=\pi(1-2M)/2$, $\epsilon(k)=(v_-^2k^2+\Delta^2)^{1/2}$,
$\Theta_s(\omega)$ is a unit step function,
and $c_p^{z,\perp}$ are positive numerical constants.
The $\delta$-function peak in Eq.\ (\ref{Sqw_nematic_FM1}) will
have a finite width when the nonlinearity of the
low-energy dispersion is included.
In the SDW$_2$ phase with AF $J_1>0$,
$k_2$ in Eq.~(\ref{Sqw_nematic_FM2}) should be replaced with
$\tilde k_2=\pi(1+2M)/2$.
The longitudinal part $S^{zz}(k,\omega)$ in the higher multipolar and
SDW$_p$ phases ($p\ge 2$) is also obtained
from Eq.~(\ref{correlators_boson_1}) as
\begin{subequations}
\label{Sqw_multipolar}
\begin{eqnarray}
\label{Sqw_multipolar_1}
&&
S^{zz}(k\sim 0,\omega) = p^2K|k|\delta(\omega-v|k|),\\
\label{Sqw_multipolar_2}
&&
S^{zz}(k\sim \pm k_p,\omega) =
\frac{c_p^z\Theta_s(\omega-v|k\mp k_p|)}{[\omega^2+v^2(k\mp k_p)^2]^{1-K}},
\end{eqnarray}
\end{subequations}
where $k_p=\pi(1-2M)/p$.
For comparison,
$S^{\alpha\beta}(k,\omega)$ in the standard TL liquids,
e.g., spin-$\frac{1}{2}$ AF Heisenberg chains
under a magnetic field, have the form~\cite{Gia_text}
\begin{subequations}
\label{Sqw_chain}
\begin{align}
\label{Sqw_chain_1}
S^{zz}(k\sim 0,\omega) =&\, K|k|\delta(\omega-v|k|),\\
\label{Sqw_chain_2}
S^{zz}(k\sim \pm k_1,\omega) =&\,
\frac{c^z_1\Theta_s(\omega-v|k\mp k_1|)}
{[\omega^2+v^2(k\mp k_1)^2]^{1-K}},\\
\label{Sqw_chain_3}
S^{+-}(k\sim \pm \pi,\omega) =&
\frac{c^\perp_1\Theta_s(\omega-v|k\mp\pi|)}
{[\omega^2+v^2(k\mp\pi)^2]^{1-1/(4K)}},
\\
\label{Sqw_chain_4}
S^{+-}(k\sim \pm 2\pi M,\omega) =&\,
\frac{\tilde c_1^\perp\Theta_s(\omega-v|k\mp 2\pi M|)}
     {[\omega\mp v(k\mp2\pi M)]^2}
\nonumber\\
&
\times
\left[\omega^2-v^2(k\mp2\pi M)^2\right]^\gamma,
\end{align}
\end{subequations}
where $\gamma=K+\frac{1}{4K}$,
and $\tilde c_1^\perp$ is a positive constant.

\begin{figure}
\begin{center}
\includegraphics[width=7cm]{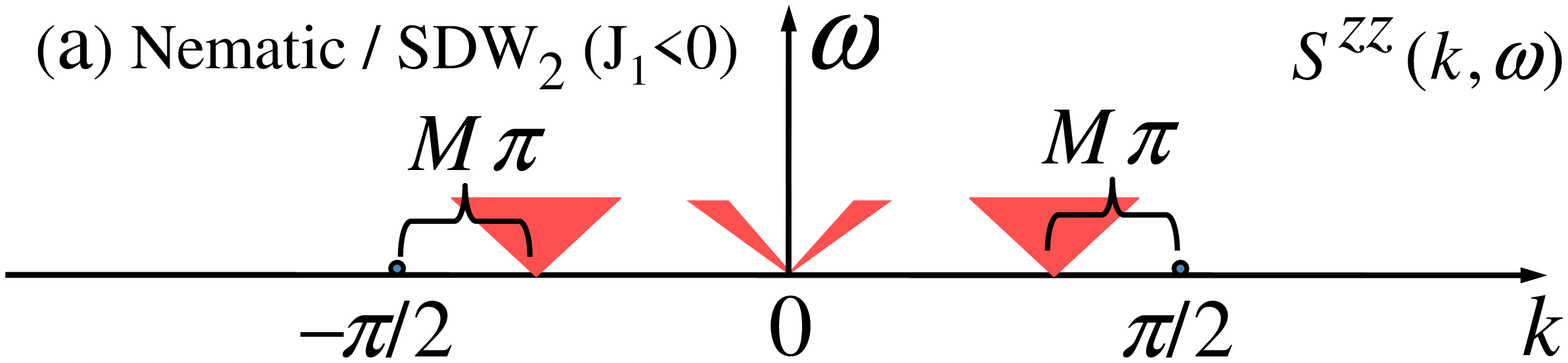}\\
\includegraphics[width=7cm]{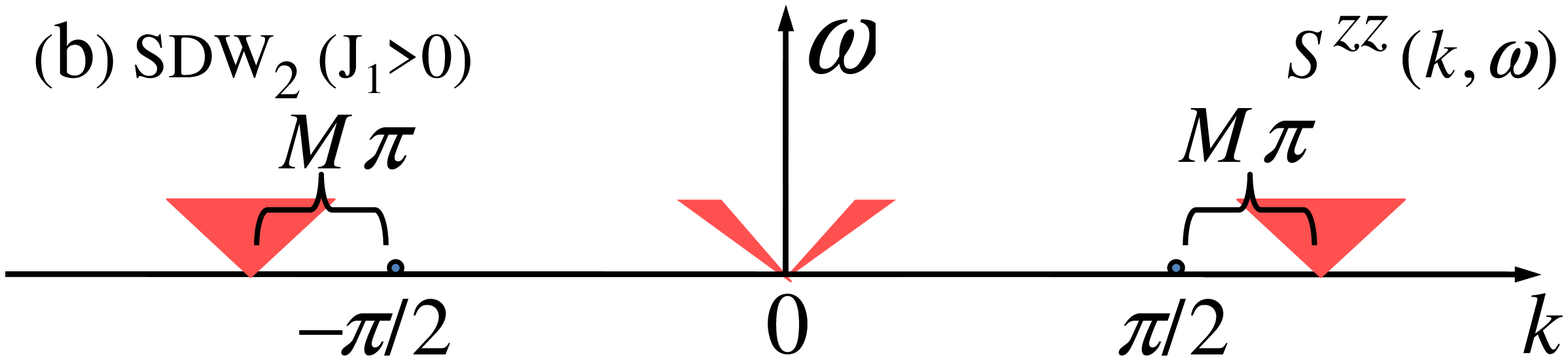}\\
\includegraphics[width=7cm]{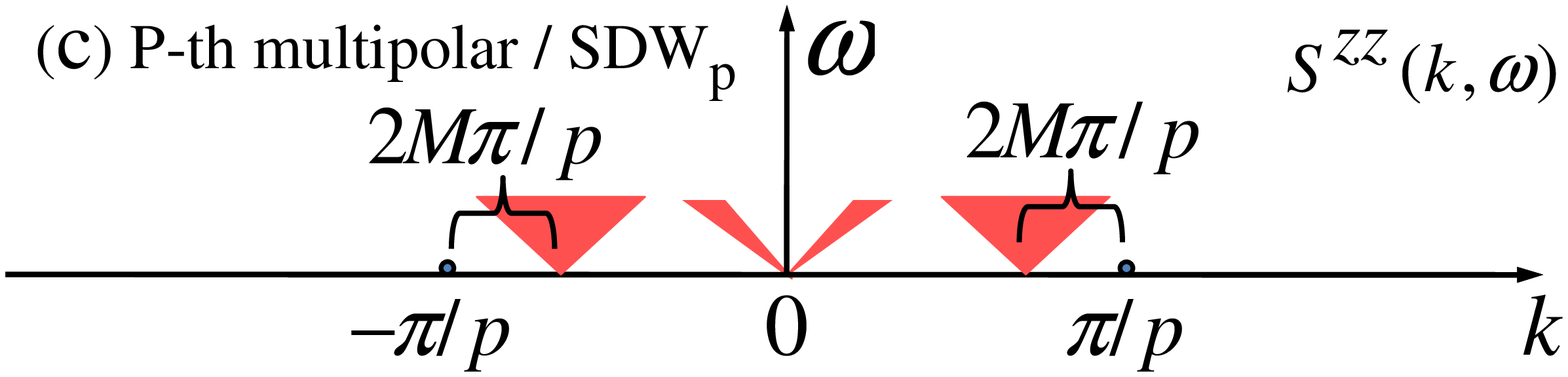}\\
\includegraphics[width=7cm]{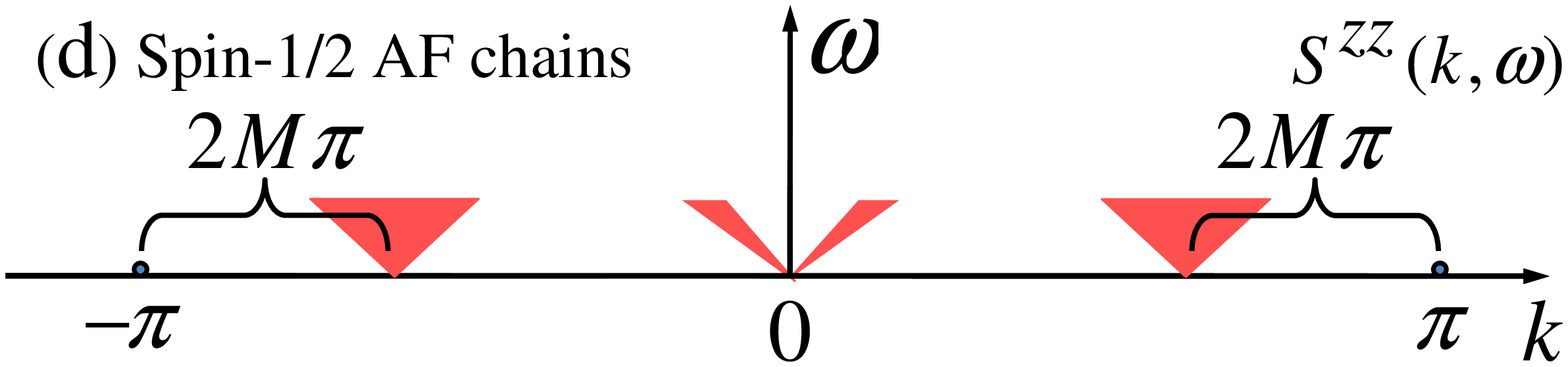}\\
\includegraphics[width=7cm]{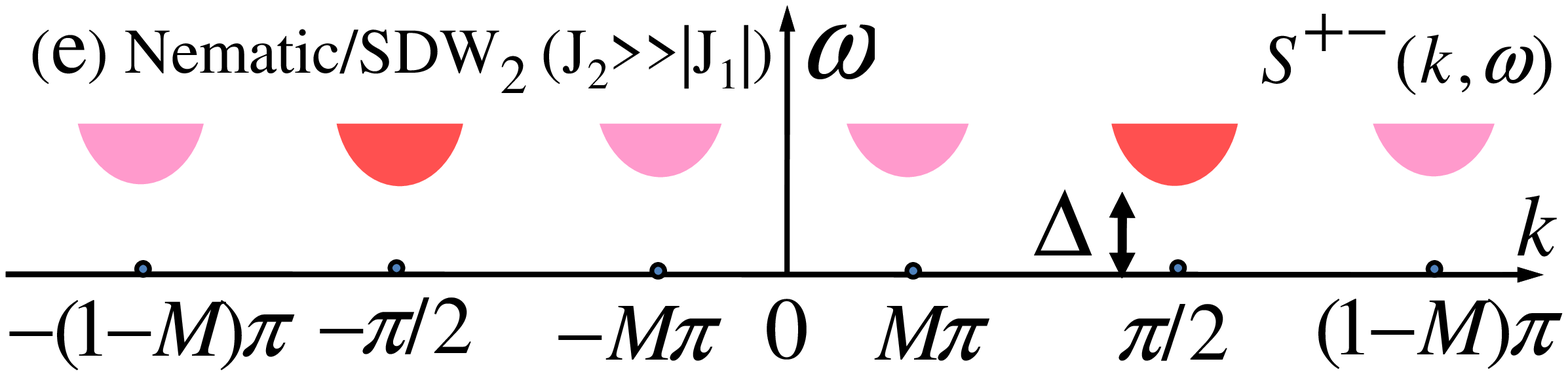}\\
\includegraphics[width=7cm]{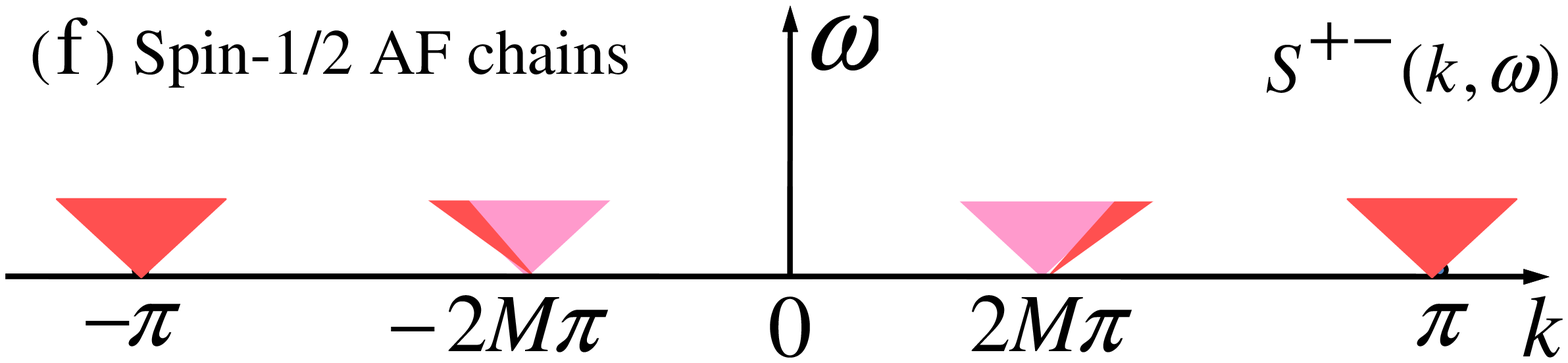}
\end{center}
\caption{(color online)
Low-energy relevant parts of $S^{zz}(k,\omega)$ and $S^{+-}(k,\omega)$
in the multipolar and SDW$_p$ phases of the spin-$\frac{1}{2}$ frustrated
zigzag chain in magnetic field [(a), (b), (c), and (e)] and
in the TL liquid of
spin-$\frac{1}{2}$ AF chains in magnetic field [(d) and (f)].
}
\label{Sqw_Fig}
\end{figure}

These results are depicted in Fig.~\ref{Sqw_Fig}.
The gapless excitations giving
dominant contribution to $S^{zz}(k,\omega)$
in the $p$th multipolar or the
SDW$_p$ TL liquid are located
at $k=\pm(1-2M)\pi/p$, when $J_1<0$.
These wave numbers,
inversely proportional to the
number $p$ of magnons forming a bound state,
are equal to the ``$2k_F$'' of the hard-core Bose liquid
of bound $p$ magnons\cite{Hikihara1} (note that fermions and hard-core 
bosons are equivalent in one dimension).
The result for the ordinary TL liquid (e.g., the AF Heisenberg chain)
corresponds to
the case $p=1$, or the limit $J_1\to0$
(the lattice unit equals two in this case).
Furthermore, one can discriminate between
the SDW$_2$ phases in $J_1<0$ and in $J_1>0$
by observing the shift of the gapless points from $k=\pi/2$
in $S^{zz}(k,\omega)$.
Another manifest difference
between multipolar and ordinary TL liquids
is that the transverse
component $S^{+-}(k,\omega)$ has a gap in the multipolar phases,
while that of the ordinary TL liquids is gapless.
These features in
$S^{\alpha\beta}(k,\omega)$ can be employed as definite
signatures of the multipolar and the SDW$_p$ phases
in the model~(\ref{zigzag}).

To conclude, we have studied dynamical response of the multipolar TL liquids
in the spin-$\frac{1}{2}$ frustrated zigzag chains in applied magnetic field.
The NMR relaxation rate $1/T_1$ in the multipolar TL liquids
shows algebraic decay with lowering temperature, which is distinct from
the diverging behavior in conventional TL liquids like
the spin-$\frac{1}{2}$ AF chains
(see Fig.~\ref{NMR_rate}).
Furthermore the wave-number and the magnetization
dependence of the gapless modes in
the dynamical structure factors $S^{\alpha\beta}(k,\omega)$
can provide us with clear evidence
for the presence of the multipolar liquids as well as SDW$_p$ regions
of bound magnons.
Our arguments are also applicable to multipolar phases in higher-spin
chains, where $K$ can become larger than unity due to soft-core
repulsion of bosons.


We thank T. Hikihara and M. Takigawa
for stimulating discussions.
This work was supported by Grants-in-Aid for Scientific Research
from MEXT, Japan (Grants No.\ 17071011, No.\ 20046016, and No.\ 16GS0219).

\end{document}